\documentclass[a4paper,conference]{IEEEtran}
\IEEEoverridecommandlockouts

\usepackage{cite}
\usepackage{amsmath,amssymb,amsfonts}
\usepackage{algorithmic}
\usepackage{graphicx}
\usepackage{textcomp}
\usepackage{xcolor}
\usepackage{subcaption}
\usepackage{float}
\usepackage{enumitem}
\usepackage{pbalance}
\usepackage{xurl}

\def\BibTeX{{\rm B\kern-.05em{\sc i\kern-.025em b}\kern-.08em
    T\kern-.1667em\lower.7ex\hbox{E}\kern-.125emX}}
\begin{document}

\title{\huge Sandbox-Enabled Digital Twin for Cyber-Physical Systems 

\thanks{
$\dagger$ Equal contribution.\\ This work is supported in part by DOE NETL grants DE-CR0000051 and DE-CR0000017, DOE GAANN Fellowship, and Center for AI and Robotics, funded by Tamkeen
under the NYUAD Research Institute Grant CG010.
}
}

\author{\IEEEauthorblockN{Meet Udeshi$^\dagger$, Md Raz$^\dagger$, Prashanth Krishnamurthy, Ramesh Karri, Farshad Khorrami}
\IEEEauthorblockA{\textit{Dept. of Electrical and Computer Engineering, NYU Tandon School of Engineering, Brooklyn, NY 11201, USA}\\
\texttt{\{m.udeshi,md.raz,prashanth.krishnamurthy,rkarri,khorrami\}@nyu.edu}}
}

\maketitle

\begin{abstract}
    Firmware/software in cyber-physical system (CPS) embedded devices/controllers can have vulnerabilities stemming from multiple sources such as weak security practices, outdated libraries, or supply chain attacks that induce adversarial effects under plant state-based triggers.
    However, pre-deployment validation of CPS controllers typically relies on digital twins that model controller logic as a black box. On the other hand, side channel monitoring and anomaly detection of CPS controller firmware/software is complementary, but is typically exercised with synthetic inputs or under specific CPS operational profiles and does not simultaneously track software execution and CPS plant evolution.
    To bridge this gap, we present a closed-loop digital twin framework that hosts unmodified controller binaries in a Linux sandbox (SaMOSA) with its I/O rerouted to an external plant simulator.
    The framework captures four time-synchronized side channels (hardware performance counters, system calls, disk activity, network activity) alongside plant state and provides orchestration hooks for automated, repeatable, parameterized runs.
    We demonstrate the framework on an OpenPLC runtime controlling a Modbus-connected IEEE 14-bus power system, and also briefly discuss application to robotics systems.
    The synchronized traces correlate internal controller execution with plant events, providing an observability foundation for online testing, coverage analysis, and vulnerability detection.
\end{abstract}

\begin{IEEEkeywords}
Cyber-physical systems, digital twin, side channels, sandbox, QEMU emulation, anomaly detection
\end{IEEEkeywords}


    \begin{figure*}[ht]
        \centering
        \includegraphics[width=1\linewidth]{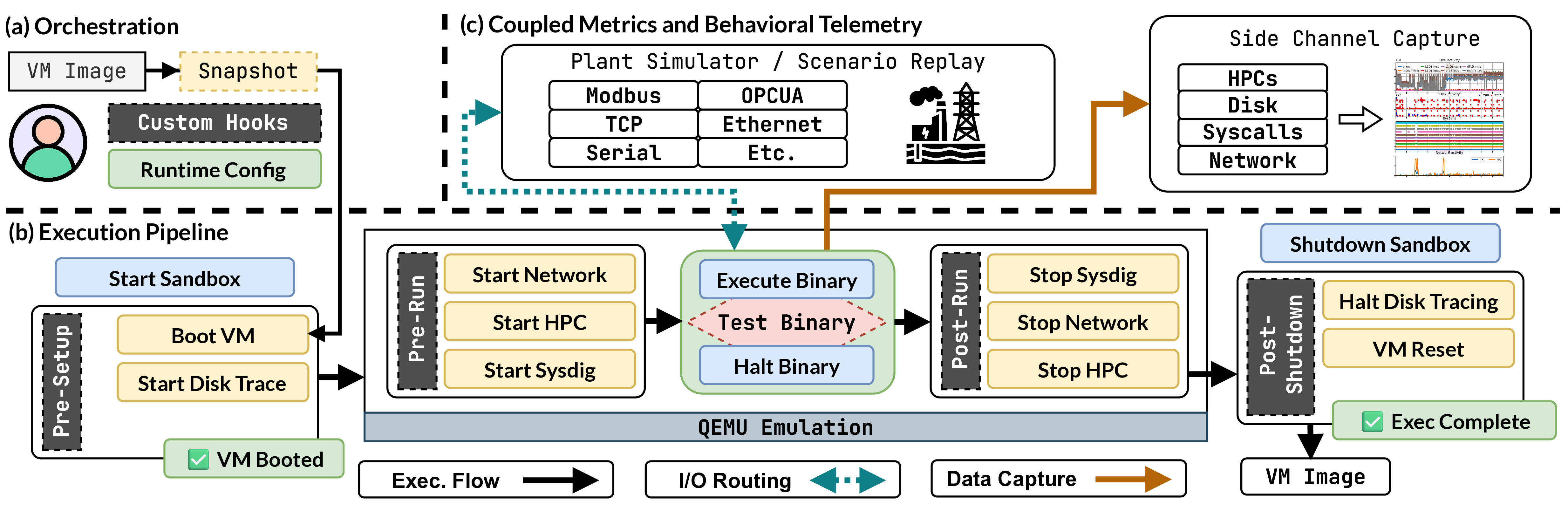}
        \vspace*{-0.15in}
        \caption{Orchestration and execution framework for the closed-loop digital twin sandbox, including flow, I/O routing, and hooks. }
        \label{fig:arch}
        \vspace*{-0.15in}
    \end{figure*}

\section{Introduction and Background}

    \subsection{Motivation}

    Industrial control systems and other cyber-physical systems (CPS) are increasingly targeted by adversaries~\cite{khorrami2016cybersecurity}, making it crucial to analyze firmware/software in CPS embedded devices/controllers to detect vulnerabilities. Vulnerabilities can stem from multiple sources, such as use of weak security practices or outdated libraries or might be intentionally injected in supply chain attacks. The need for robust vulnerability analysis is highlighted by attacks such as Stuxnet~\cite{kerr2010stuxnet}, Triton~\cite{michail2021vulnerabilities}, and Kinsing malware~ \cite{Kinsing1,Kinsing2} as well as other logic and runtime security issues~\cite{abbasi2016ghost,spenneberg2016plc, govil2017ladder}, which show that control logic and embedded devices can be compromised at scale. Despite these risks, controller firmware is rarely validated against the full spectrum of field-side plant conditions or runtime anomalies~\cite{costin2014large}. Pre-deployment validation typically relies on hardware-in-the-loop setups or vendor/domain-specific simulators that exercise I/O behaviors~\cite{feng2020scalable}, but offer minimal visibility into internal execution~\cite{kim2020firmae}. Gaining deeper visibility traditionally requires compiling binaries with custom instrumentation code, which is often infeasible for resource-constrained embedded systems and closed-source commercial software~\cite{preusser2022everything, wright2021challenges}.
        
    While digital twins enable plant-side analysis by allowing operators to replay incidents, sweep parameter spaces, and stress-test control logic under simulated faults~\cite{eckhart2019digital}, most implementations treat the controller as a black box and do not address internal firmware execution. Hence, vulnerabilities remain undetected unless they manifest as plant-visible signals. On the other hand, side channel or time series monitoring has been shown to effectively detect malware/anomalies on embedded and Linux-based systems~\cite{wang2016malicious,krishnamurthy2020anomaly,pearce2022detecting,udeshi2025samosa,krishnamurthy2023multimodal,krishnamurthy2024muddle,khorrami2025realtime,udeshi2024tamper,krishnamurthy2024tracking}. However, these methods are typically evaluated on physical devices or sandboxed environments under synthetic inputs, replayed traces, or specific CPS operational profiles, providing limited visibility into behavioral and side channel variations under dynamic excitations. Moreover, these methods do not track controller execution side channels simultaneously with the plant state time series, nor do they enable coordinated fuzzing/triggers across both the cyber (controller) and physical (plant) domains.

    Prior frameworks do not jointly (i)~host an unmodified controller binary, (ii)~drive it with realistic closed-loop inputs from a plant simulator with configurable fuzzing of both the controller execution and plant dynamics, (iii)~capture time-synchronized side channels from the controller alongside plant state~\cite{zaddach2014avatar,chen2016towards,clements2020halucinator}.
        To address these limitations, we present an integrated framework built atop the SaMOSA Linux sandbox~\cite{udeshi2025samosa} and make three key contributions:
        1. A digital twin framework that closes the loop between an unmodified controller binary executing in-sandbox and an external plant simulator via QEMU-native I/O rerouting.
        2. Automation of closed-loop experiments with fault-injection diagnostics and scenario replay by leveraging orchestration and side channel capture.
        3. A detailed case study of a digital twin of a programmable logic controller (PLC) running OpenPLC~\cite{alves2014openplc,autonomy_openplc_editor} which executes a Structured Text control program for a Modbus-connected IEEE 14-bus power system model, yielding time-aligned plant signals and four controller side channels~\cite{modbus2012application, iec61131}. We discuss adapting this approach to a robotic system by replacing the OpenPLC controller and power system dynamic models with ROS (Robot Operating System) packages~\cite{ROS,macenski2022robot} and Isaac Sim  simulator~\cite{nvidia2026isaacsim}.

    \subsection{Sandbox-Based Instrumentation and Digital Twins} 
        Sandboxes combine isolation and observability to execute untrusted code while capturing its dynamic behavior~\cite{ormeir2019dynamic}. Unlike static inspection, modern sandboxes record side channel traces such as system calls, hardware performance counters (HPCs), network, and disk activity to reveal internal execution patterns. This makes them ideal for control systems as they expose the behavior of unmodified controller binaries while preserving native execution.
        By contrast, digital twins prioritize plant-side fidelity but typically treat the controller as a black box via hardware-in-the-loop or behavioral co-simulation~\cite{eckhart2018towards, gehrmann2020digital}. A sandbox addresses this observability gap by hosting the binary and rerouting I/O to the plant model. By collecting time-synchronized side channels, the sandbox closes the loop, providing comprehensive visibility into both the controller's internal state and the plant's physical response.
    
    \vspace*{-0.02in}
    \subsection{The SaMOSA Sandbox} 
    \vspace*{-0.02in}
        SaMOSA instantiates this idea with QEMU full-system emulation~\cite{udeshi2025samosa, qemu} using architectures common in PLCs and other embedded systems (x86-64, ARM64, and PPC64LE) and runs full Ubuntu/Debian distributions for a realistic environment. It records four time-synchronized side channels at low-interference capture points: host-side \texttt{perf} on the QEMU process for workload-wide hardware-performance behavior, in-guest \texttt{sysdig}~\cite{sysdig} for OS-wide system calls, QEMU hypervisor tracing for timestamped disk accesses with logical block addresses, and host-side \texttt{tcpdump} on the network tap interface. Sharing host timestamps, these traces enable cross-channel correlation between controller behavior and plant events. SaMOSA further exposes four orchestration hooks, Pre-Setup, Pre-Run, Post-Run, and Post-Shutdown, that automate loading, configuration, and scaling for repeated, parameterized closed-loop runs. We attach a plant simulator to its network bridge, redirecting traffic to the simulated process instead of FakeNet\cite{fakenet} used for malware analysis.

\section{Proposed Framework and Methodology} 

    \subsection{Architecture} 

        The integrated framework (Figure~\ref{fig:arch}) interposes on the controller inputs/outputs at the emulator boundary such that binaries and runtimes execute unmodified. Controller runtime is housed in the SaMOSA full-system VM under QEMU and interfaces with a model of the physical system (a simulator process on the host) that takes as inputs controller actuation commands and returns updated sensor readings closing the control loop. SaMOSA side channels (hardware performance counters, system calls, disk activity, network traffic) are captured at runtime, while orchestration hooks (Section~\ref{sec:orchestration}) assist in configuring the simulator and arming each scenario.
        
        Controller I/O is routed to a host-side plant simulator (or replay) endpoint over the network bridge without modifying the controller binary. The guest interface is exposed via \texttt{-netdev tap} and attached as a bridge peer, carrying any IP-based industrial protocol (e.g., Modbus/TCP, OPC~UA, EtherNet/IP, MQTT, DDS transport in ROS2), where our case study uses Modbus/TCP. The same principle extends to non-network controllers through QEMU's other host-backed devices: serial links via \texttt{-serial} on a \texttt{chardev} socket backend (Modbus~RTU, DNP3, or vendor byte streams) and arbitrary character devices via \texttt{virtio-serial}. QEMU's block-device tracing, already part of SaMOSA, additionally exposes the controller's in-disk persistence behavior.

    \vspace*{-0.02in}
     \subsection{Closed-Loop Orchestration}  \label{sec:orchestration}
    \vspace*{-0.02in}
        The closed loop is assembled and driven via SaMOSA's four hook points allowing scriptable, repeatable experiments. The hooks couple to execution pipeline (Figure~\ref{fig:arch}) so that simulator steps, scenario parameters, and data collection are automated.
    
            \noindent\textbf{Pre-Setup:} Host-side preparation before the VM boots, e.g., initializing simulator state or, for serial-mediated plants, opening the host socket backing the guest \texttt{chardev}. \\
            \noindent\textbf{Pre-Run:} The plant simulator is launched on the sandbox bridge and the control program is loaded with the scenario's disturbance or fault profile. Side channel monitors are  enabled and the controller runs for the selected duration, with host-side timestamps recorded at execution and halt boundaries. \\
            \noindent\textbf{Post-Run:} After the controller halts, the final simulator state and plant signal log are exported alongside the synchronized side channel bundle for time-aligned analysis. \\
            \noindent\textbf{Post-Shutdown:} After VM teardown, the simulator is stopped, the final VM image is exported, and the host environment is reset for the next scenario.
        
\section{Case Study: OpenPLC and Power Systems} 
    
    \begin{figure}[t]
        \centering
        \includegraphics[width=0.98\linewidth]{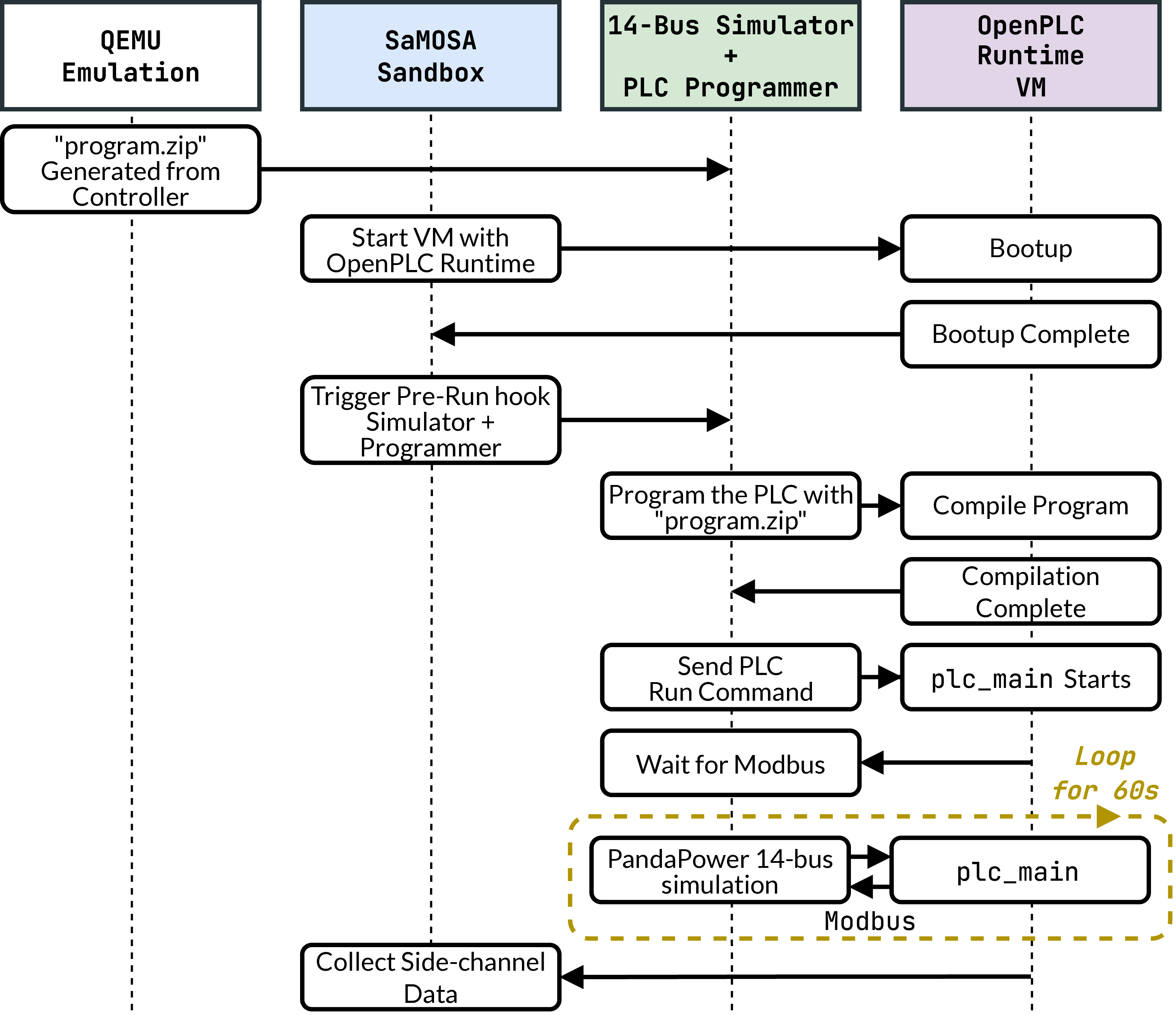}
        \caption{PLC programming and execution timeline in the case study for a power systems application.}
        \label{fig:timeline}
    \vspace*{-0.1in}
    \end{figure}
    
    We demonstrate the SaMOSA digital twin using the  IEEE~14-bus power system benchmark~\cite{pscad2022ieee14bus} controlled by an OpenPLC runtime~\cite{autonomy_openplc_editor}, hosted in an ARM64 sandbox image running Ubuntu~24.04. The sandbox emulates an ARM Cortex-A72 replicating a Raspberry Pi controller. A web API is used for OpenPLC programming and execution with I/O via Modbus and IEEE~14-bus simulator. The PandaPower Python power system flow solver package~\cite{pandapower.2018,pandapower} is used to implement the physical power system dynamics model, communicating in real-time with the sandbox-hosted OpenPLC controller over Modbus TCP to enable the closed-loop cyber-physical co-simulation. The OpenPLC implements the PI (Proportional Integral) governor controller at Generator 2 (at Bus 2) to calculate necessary adjustments to handle varying power demand due to changing load, while Bus 1 serves as the slack bus connected to the external grid. While the Generator 2 control algorithm is brought out to the sandbox-hosted OpenPLC, the remaining generators within the network are managed internally by the dynamic simulator's default dispatch and voltage regulation models. Generator 2 telemetry data (frequency, electrical power) from the Python-based simulator is sent via Modbus TCP to OpenPLC at every sampling instant upon which OpenPLC computes the PI controller algorithm and sends back a regulated mechanical power command. In the simulation, at time $t=3.0$ seconds, a 20 MW load step disturbance is applied at Bus 14 on the distant edge of the network, testing the frequency response and governor action of Generator 2 as it reacts to the remote load change. The closed-loop CPS digital twin tracks both the digital side channels from the sandbox and the physical power system time series data.

    \begin{figure}[!t]
        \centering
        \begin{subfigure}[t]{\linewidth}
        \includegraphics[width=\linewidth]{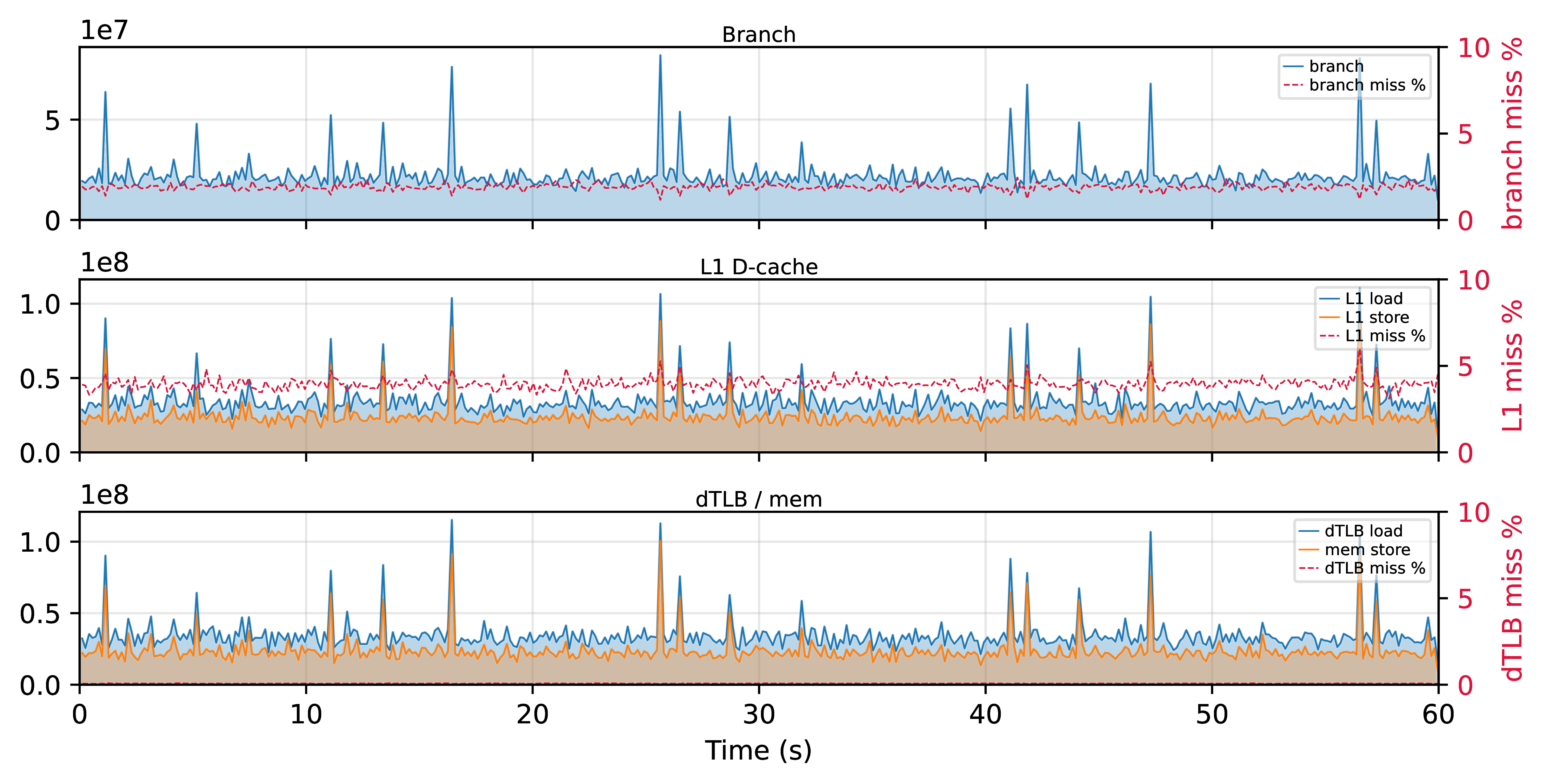}
               \vspace*{-0.3in}
            \caption{Hardware Performance Counters (HPC)}
            \label{fig:hpc}
            \vspace*{0.1in}
        \end{subfigure}
        \begin{subfigure}[t]{\linewidth}
       \includegraphics[width=\linewidth]{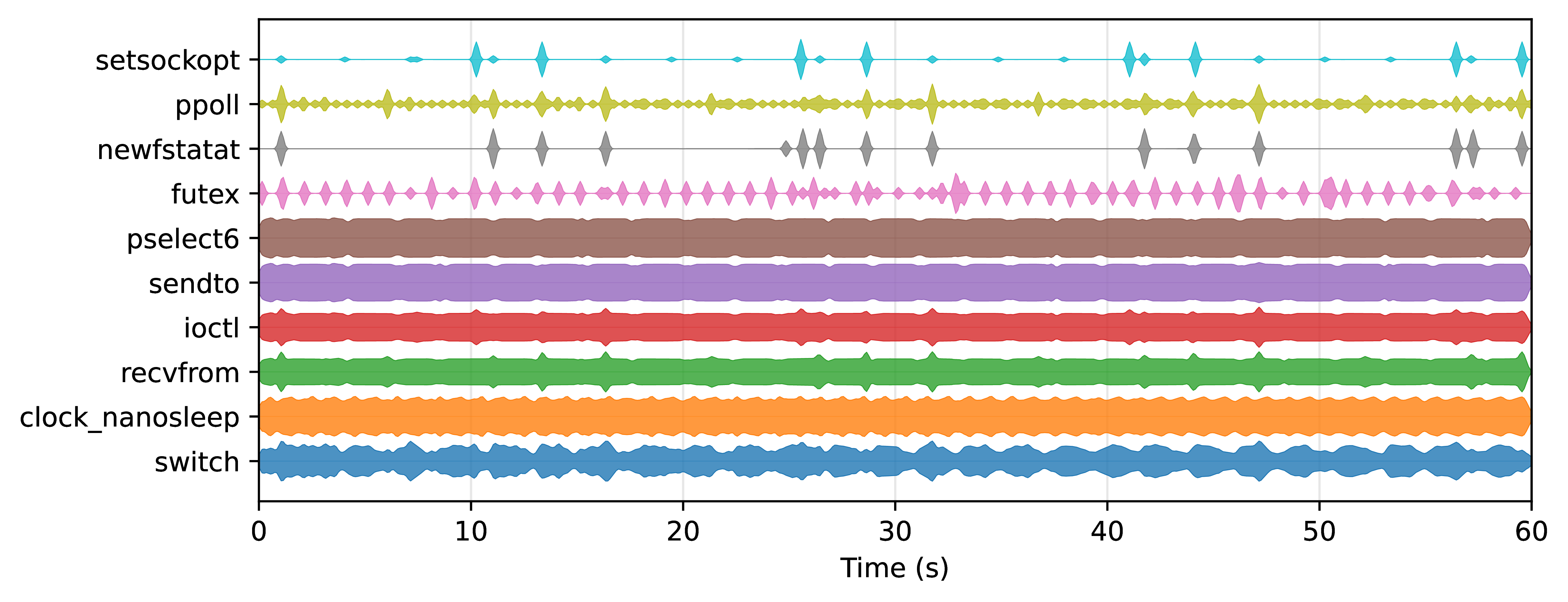}
          \vspace*{-0.3in}
            \caption{Syscalls}
            \label{fig:syscall}
            \vspace*{0.1in}
        \end{subfigure}
        \begin{subfigure}[t]{\linewidth}
        \includegraphics[width=\linewidth]{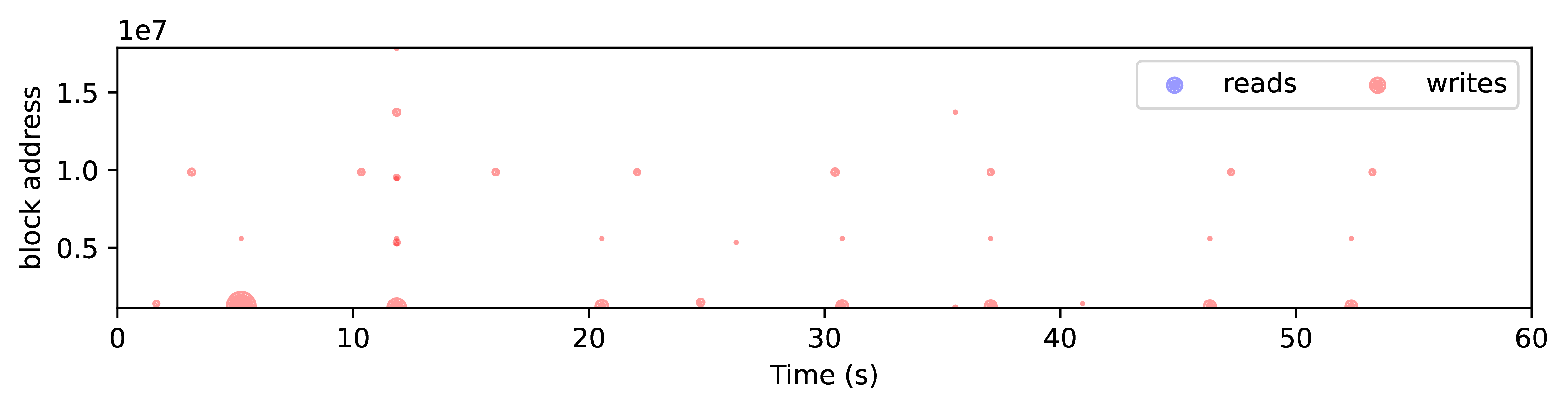}
            \vspace*{-0.3in}
            \caption{Disk access}
            \label{fig:disk}
            \vspace*{0.1in}
        \end{subfigure}
        \begin{subfigure}[t]{\linewidth}
        \includegraphics[width=\linewidth]{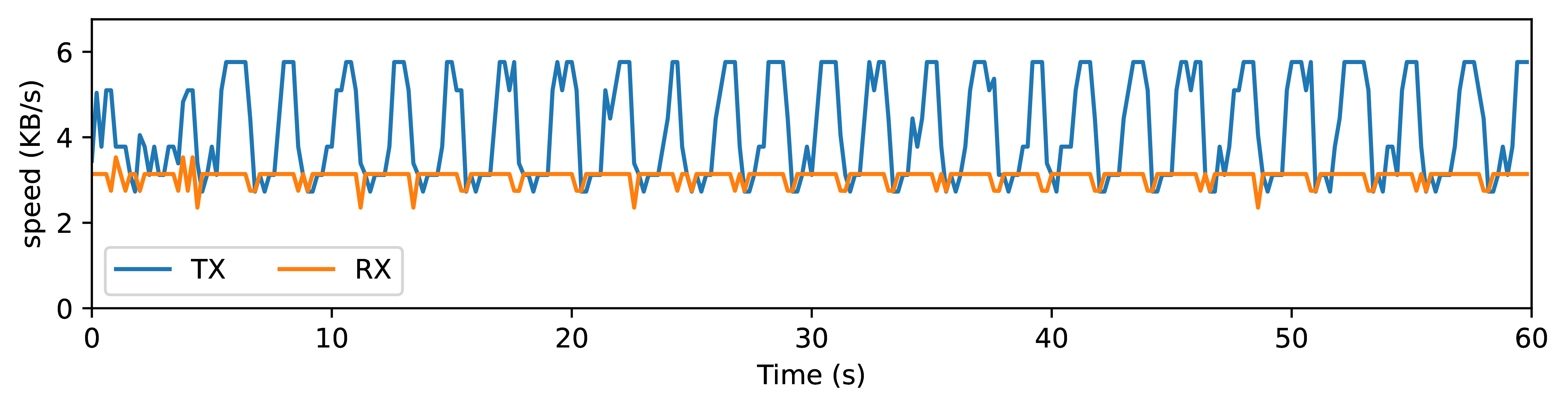}
            \vspace*{-0.3in}
            \caption{Network traffic}
            \label{fig:network}
            \vspace*{0.1in}
        \end{subfigure}
        \begin{subfigure}[t]{\linewidth}
        \includegraphics[width=\linewidth]{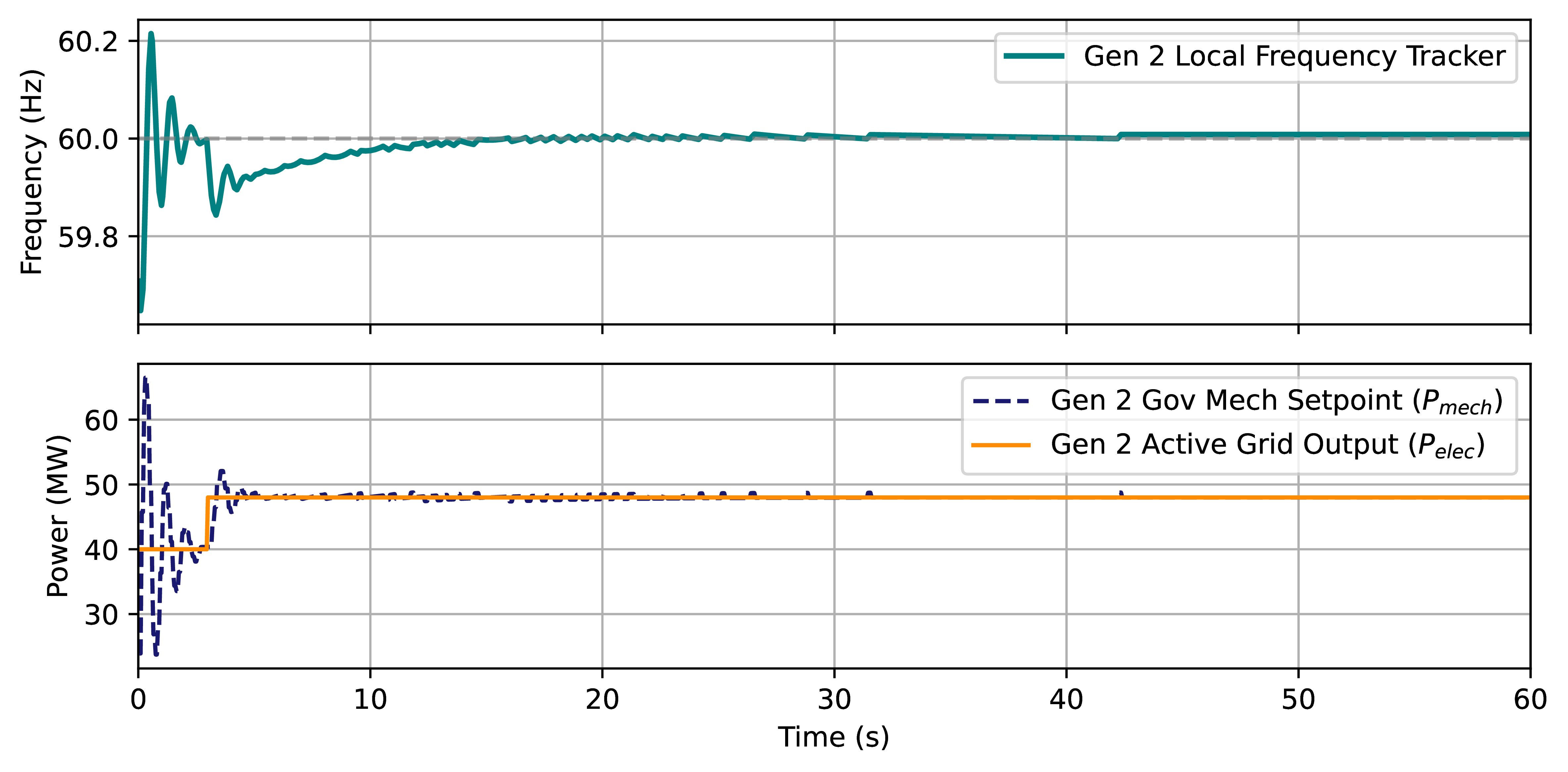}
            \vspace*{-0.3in}
            \caption{Closed-loop power system simulator signals}
            \label{fig:sim}
            \vspace*{0.1in}
        \end{subfigure}
        \caption{Side channels from SaMOSA sandbox + power system simulation output for OpenPLC IEEE 14-bus case study.}
        \label{fig:case1}
        \vspace*{-0.1in}
    \end{figure}
    
    Figure~\ref{fig:timeline} outlines the digital twin execution timeline. The 14-bus controller is implemented in Structured Text using the OpenPLC Editor, which emits a standalone \texttt{program.zip}. SaMOSA  boots the OpenPLC runtime VM and triggers the ``Pre-Run'' hooks, which start the IEEE~14-bus simulator and PLC programmer. The programmer loads \texttt{program.zip} via the web API, triggering compilation in the runtime, then calls the API to start the PLC, launching the \texttt{plc\_main} binary. The simulator, waiting for a Modbus connection, proceeds once \texttt{plc\_main} is running and a steady Modbus exchange begins. Over 60~seconds, the simulator queries the PLC for control inputs, computes the power system response, and returns outputs each time step, while the PLC executes the controller at its scan-cycle rate. SaMOSA records this execution and network activity across its four side channel types, then shuts down the VM and collects the data once the run completes.

    We plot SaMOSA's side channel data alongside simulation history to expose correlations between controller execution and the physical process. Figure~\ref{fig:hpc} shows the HPC counters, with subplots for branches and branch-miss rate; L1 loads, stores, and miss rate; and dTLB loads, memory stores, and dTLB miss rate. The counters stay consistent apart from occasional spikes; these do not align with physical events, suggesting they reflect operating-system activity rather than the control task. Figure~\ref{fig:syscall} shows the 20 most frequent system calls as per-window violin plots. The pattern is periodic at the scan-cycle rate, dominated by \texttt{pselect6}, \texttt{sendto}, \texttt{recvfrom}, \texttt{ioctl}, and \texttt{clock\_nanosleep}, and exhibits the network I/O and sleep-driven context switching expected of a scan cycle that reads and writes Modbus values, executes the program, and sleeps until the next cycle. Figure~\ref{fig:disk} plots disk activity as a bubble scatter and shows few writes and absent reads indicative of the runtime executing almost entirely from memory. Figure~\ref{fig:network} shows transmit and receive speeds, both periodic every two seconds. Figure~\ref{fig:sim} plots Generator 2 frequency and power signals from the power system simulator, showing the response to a load step at 3~seconds.

    SaMOSA captures rich, multi-modal side channel data from an OpenPLC runtime performing a realistic power system control task. The syscall and network channels recover the periodic scan cycle and its Modbus I/O, the near-absent disk activity indicates in-memory execution, and the otherwise-stable HPC counters isolate OS-level interference from control activity, highlighting a key nuance in interpreting hardware-level telemetry.
    Together, the side channel signals captured by SaMOSA provide a detailed picture of PLC behavior that can be correlated with the underlying physical simulation.

\section{Application to Robotic Systems} 

    \begin{figure}[!t]
        \centering
        \includegraphics[width=1\linewidth]{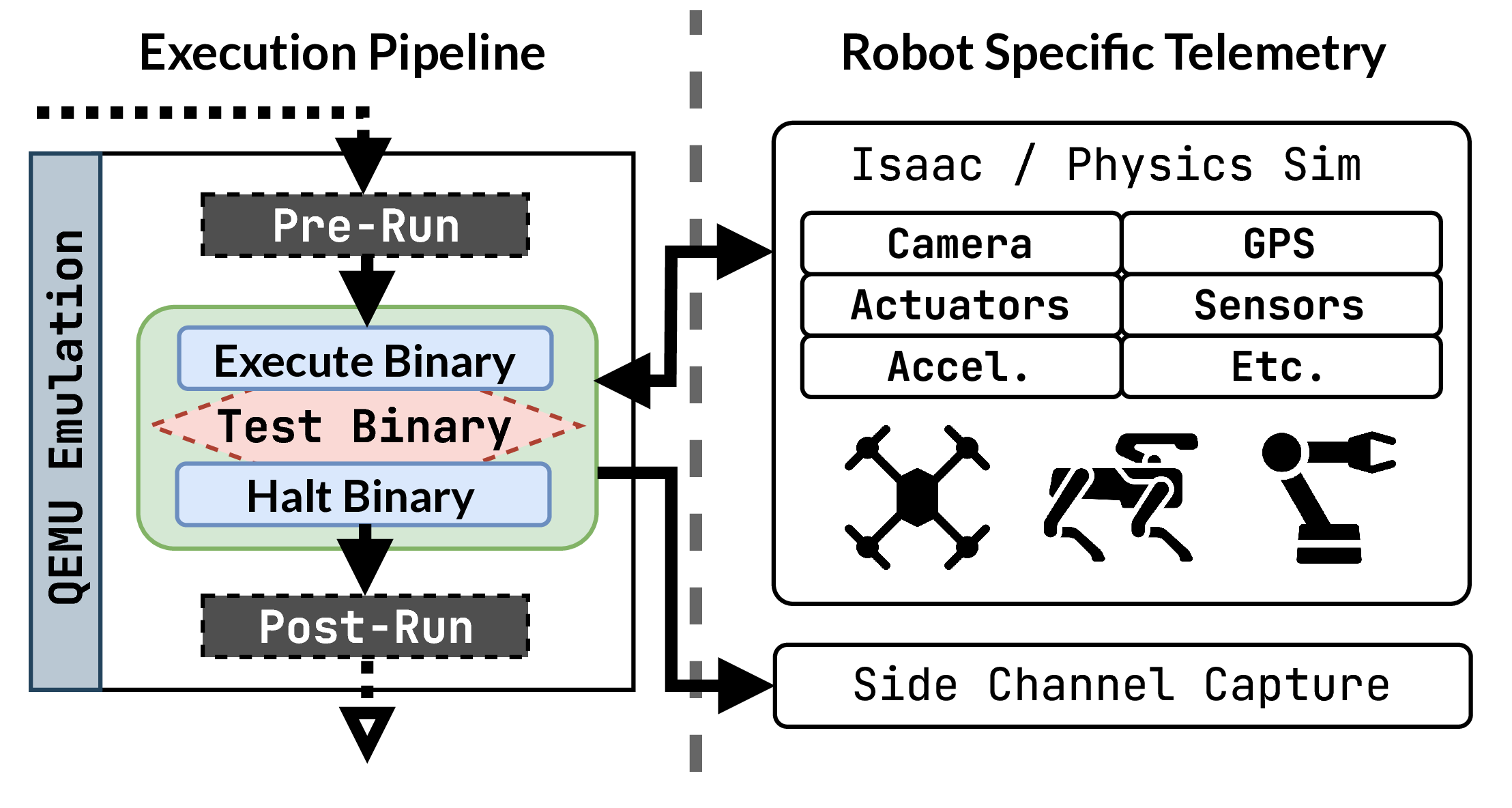}
        \caption{Adapting the digital twin framework and execution pipeline to a robotics application.}
        \label{fig:robot}
        \vspace*{-0.05in}
    \end{figure}
    
    This digital twin framework is general. While the sandbox and orchestration mechanisms are scalable to different application domains, the simulator and coupled I/O modalities are domain-specific. In robotics, ROS~\cite{ROS,macenski2022robot} and the flight or locomotion controller stack atop it can be hosted in SaMOSA 
    and its transport rerouted over the network bridge to a robot physics simulator, Isaac~Sim~\cite{nvidia2026isaacsim}. Isaac~Sim replaces the PandaPower plant and models the robot and its environment, taking as inputs actuation commands (joint torques or velocity setpoints) and returning the sensor data stream (Figure~\ref{fig:robot}).
    Distributed ROS systems involving multiple devices (e.g., a distributed sensor fusion system or a controller coordinating multiple drones or multiple robotic arms) can likewise be modeled by launching multiple instances of SaMOSA, one per device.
    The coupled interfaces change when compared to the PLC case study as a robot's perception and state-estimation inputs span many modalities including inertial measurements, RGB and depth cameras, GNSS/GPS, time-of-flight and LiDAR ranging, joint encoders, and force/torque or foot-contact sensors, each of which the simulator can generate, the orchestration layer can perturb, or a captured trace can replay. Passing through the same rerouting boundary, each becomes an injection point for controlled experiments without touching the binary, such as: \\
        \noindent\textbf{Environmental scenarios:} terrain, obstacle, wind, or lighting changes that stress perception and control, with back-to-back replays for A/B comparison. \\
        \noindent\textbf{Sensor faults and spoofing:} dropout, bias, noise, stuck values, GPS spoofing, or sensor artifacts injected per modality to study fault tolerance and detection. \\
        \noindent\textbf{Actuator faults:} motor loss, joint lock, or reduced torque to probe degraded behavior. \\
        \noindent\textbf{Configuration and algorithm testing:} swapping ROS parameters, QoS settings, or
        estimator/planner/controller implementations under identical replay conditions.
    
The side channels are captured alongside time series of simulated plant states, coupling the controller stack's internal behavior to the robot's physical responses. The online-testing, coverage, and anomaly-detection goals are analogous to the PLC case, but extended to the richer interfaces and modalities in the robotics context~\cite{chung2019smartmalware, witte2018consistency}. Further development of the robotics applications of the proposed framework with integration of high-bandwidth I/O modalities such as camera and LiDAR and sim-time synchronization implementations for robotics simulators will be addressed in future work.

\section{Conclusion} 

    We presented a closed-loop digital twin framework that pairs the SaMOSA Linux sandbox with an external plant simulator, hosting an unmodified controller binary and rerouting its I/O at the QEMU device boundary so that multiple time-synchronized side channels can be captured alongside plant state. 
    Because the orchestration hooks make runs scriptable and repeatable, the framework is a foundation for larger studies, including fault-injection sweeps, coverage analysis, side channel anomaly detection, and vulnerability analysis.

\newpage

\bibliographystyle{IEEEtran}
\bibliography{refs}

\end{document}